\documentclass{pasj00}

\begin{document}

\SetRunningHead{}{}
\Received{}
\Accepted{}

\title{Long-term X-ray variability of quasars in the Lockman Hole field
	observed with ROSAT}

\author{Yu-ichiro \textsc{Ezoe},\altaffilmark{1}
        Naoko \textsc{Iyomoto},\altaffilmark{2}
        Kazuo \textsc{Makishima},\altaffilmark{1,3}\\
        and\\
        G$\ddot{\rm u}$nter \textsc{Hasinger} \altaffilmark{4}
        }

\altaffiltext{1}{Department of Physics, University of Tokyo,
         7-3-1 Hongo, Bunkyo-ku, Tokyo 113-0033, Japan}
\altaffiltext{2}{The Institute of Space and Astronautical Science,
     3-1-1 Yoshinodai, Sagamihara, Kanagawa 229-8510, Japan}
\altaffiltext{3}{The Institute of Physical and Chemical Research (RIKEN),
       2-1 Hirosawa, Wakho-si, Saitama 351-0198, Japan}
\altaffiltext{4}{Max-Planck-Institut f$\ddot{u}$r Extraterrestrische Physik,
       85740 Garching, Germany}

\maketitle

\begin{abstract}
An improved method is utilized to estimate the X-ray power spectral
densities (PSD) and the variation time scales of three quasars in
the Lockman Hole field. Five archival ROSAT PSPC data covering two
year range are analyzed. To estimate PSD from sparse and unevenly-sampled
lightcurves, a forward-method approach with extensive Monte-Carlo simulations
is adopted. A broken power-law type PSD with a constant Poisson noise
component is assumed with a break frequency $f_{\rm b}$.
Then, assuming the PSD slope $\alpha$ as $-2<\alpha<-1$, $1/ f_{\rm b}$
is constrained as $\gtrsim$ 25 days for one object,
while the constraints on the other two objects are very weak.
The long time scale of the one object is consistent with the view
that luminous AGNs host massive black holes.
\end{abstract}

\section{Introduction \label{intro}}


Massive black holes (BHs) have been considered as the
central engine of the active galactic nuclei (AGNs). 
In fact, utilizing the stellar and gaseous kinematics,
many dark mass concentrations, probably BHs, have been detected
at the nuclei of many nearby active galaxies
(e.g., Kormendy, Richstone 1995; Miyoshi et al. 1995;
 Gebhardt et al. 2001). However, these methods are no longer valid
for more distant AGNs, including in particular quasars (QSOs),
because of obvious technical difficulties such
as smaller angular scales and lower surface brightness
of the stellar light, as compared to nearby objects.


There are alternative ways of estimating the mass of the central BH.
One is the reverberation mapping method (Peterson 1993; Netzer,
Peterson 1997), and another is to utilize the random intensity
variability of AGNs. The latter method has been utilized over a wide
wave length range from radio to X-rays and $\gamma$-rays (e.g., Krolik
et al. 1991; Edelson et al. 1996), although the exact origin of such a
variability is still unclear. Like the reverberation mapping method, it
can be applied to both nearby and distant AGNs. In case of Seyfert galaxies,
luminous X-ray sources are generally more variable on long time scales
than on short time scales (e.g., Nandra et al. 1997; Markowitz, Edelson
2001). This property is consistent with the idea that the variation
time scale is proportional to the size of the emitting region in AGNs, and
hence to the BH mass. As an indicator of the time scale of variability, break
frequency in the power spectrum density (PSD) in the X-ray band is frequently
utilized. For example, Edelson, Nandra (1999) and Chiang et al. (2000)
estimated the BH masses of NGC 3516 and NGC 5548, respectively, assuming that
the break frequency is proportional to the BH mass, and determining the
coefficient of proportionality referring to Cyg X-1.
The estimated BH masses are consistent with those from the kinematics 
or the reverberation mapping techniques, as long as the latter is available. 
These results indicate that we can utilize the intrinsic X-ray variability
time scale of AGNs as a BH mass indicator.


High-luminosity AGNs, including QSOs in particular, have been found to have
such a long time scale of variability, up to a few years, that the analysis of
their behavior needs long observations. Such long observations in X-rays are
generally limited to all-sky monitoring of the brightest objects. Moreover,
even if there are observations spanning over several years, such data usually
suffer from window function (Fourier transform of the observational sampling)
convolved with the true PSD of the source variation. Therefore, it is usually
very difficult to reliably estimate the PSDs of QSOs over low frequency ranges.


We have developed a method of estimating the AGN variability time scales from
sparse and unevenly sampled lightcurves, utilizing structure function (SF) and
extensive Monte Carlo simulations (Iyomoto 1999). As described by Simonetti et
al. (1985), the SF is mathematically equivalent to the PSD, but less affected
by data gaps. Assuming various PSDs, we generate Monte Carlo lightcurves,
which are then subjected to the sampling window of the actual observations.
We convert them into SFs and compare with the observed SFs.
This method is similar to the ``response method'' developed by
Done et al. (1992) and Green et al. (1999), although their method utilizes
PSD instead of SF.
Employing our method, we analyzed long-term X-ray lightcurves
of a few objects observed with ASCA, assuming a broken power-law type PSD
(Iyomoto, Makishima 2001; Ezoe et al. 2001). Iyomoto, Makishima (2001)
constrained the break frequency $f_{\rm b}$ of the PSD of the M81 nucleus,
a low luminosity AGN, as 1/$f_{\rm b}>$ 800 days. They hence estimated the BH
mass of the source to be $>$ $4\times10^7M_\odot$, assuming the mass-to-time
scale proportionality.


In this paper, we investigate the long-term variability of three QSOs
in the Lockman Hole field, utilizing all the archival PSPC data
of the ROSAT Deep Survey spanning over two years.

\section{Observations and Source Selection\label{obs}}


The Lockman Hole field was observed five times with the ROSAT PSPC
in the period of 1991-1993. Table \ref{obslog} gives a summary of these
observations. The five pointings are all centered at the J2000 position
(10$^{\rm h}$52$^{\rm m}$, 57$^{\circ}$21$^{\prime}$36$^{\prime\prime}$).
We utilize the archival event files with the default screening,
achieving a total exposure time of 203 ksec. Studies of sources in this
field have been published in several papers; the detection of X-ray sources
and their properties are given in Hasinger et al. (1998; hereafter Paper I),
the optical identification is reported in Schmidt et al. (1998; Paper II) and
Lehmann et al. (2001; Paper III), while the radio identification is described
in de Ruiter et al. (1997; Paper IV).


Among point sources in the central 40$^{\prime}$ field-of-view of the PSPC,
we selected objects for our timing analysis through the following four steps.
First, we selected sources which are optically identified with AGNs 
(Papers II and III). Then, among them, we have discarded those
which are confused with other sources within the position resolution of the 
PSPC. Third, for each chosen source, we accumulated photons within
45$^{\prime\prime}$ of its X-ray center, and then selected only bright sources
with count rate $\gtrsim 0.01$ ct s$^{-1}$, corresponding to signal to noise
ratio $\gtrsim 5$ when photons are integrated within 1-day. 
The background photons are accumulated from a blank sky 
region of the same PSPC image. Finally, in order to clearly limit their
variation time scales, we made lightcurves of the sources and discarded those
whose root-mean-square (rms) variations over the five observations are less
than $\sim$ 100\% of the average of their photon-counting errors
(Poisson errors).
Through these criteria, we have selected three sources, No.28, No.32 and
No. 37, according to the nomenclature in Paper I. Table \ref{Xrayprop}
summarizes their basic properties, where we also give the ratio of
their Poisson errors to their rms variations. The method of the lightcurve
analysis are described in the next section. Below we summarize the optical
and radio information of these sources.


In the optical wavelength range, the three sources are identified with AGNs
located at redshifts $z>0.1$ (Paper II), and hence, can be classified as QSOs
according to the conventional classification with $z$. Their strong emission
lines (Paper III) rule out their blazer interpretation. Two of them, sources
No.32 and No.37, are identified with broad line type I AGNs,
while source No.28 is identified with a narrow line type II AGN.
Sources No.28 and 32 have radio
counterparts at 1465 and 1515 MHz, as shown in table \ref{Xrayprop}, while 
source No.37 does not, within the sensitivity limit of $\sim$
0.12 mJy (Paper IV). Because the optical to radio flux ratio, log $f_{\rm 5GHz}
$/$f_{\rm B}$, become $\sim$2 for the two radio-identified sources, assuming
$f_{\rm 1.5GHz} \cong f_{\rm 5GHz}$, they can be classified as radio loud QSOs
(RLQs). Thus, source No.28 is considered as a candidate for a type II QSO with
relatively strong radio emission, sources No.32 as a RLQ, and No.37 as a
candidate for a radio quiet QSO (RQQ).

\section{Lightcurves \label{lc}}


We made lightcurves of the three sources in the 0.1--2.4 keV band for each
observation, following the scheme of the ROSAT standard analysis. We also
made the background lightcurves, and subtracted it. Because response files
changed between the first observation and the others, we must be cautious about
possible changes of the effective area when comparing count rates in different
observations. Therefore we compared the effective areas as a function of
energy, utilizing these two response files and ancillary response function
files at the position of the individual sources. We have confirmed that this
effect is at most $\sim$ 10\% in this energy range,  which is within the
typical
photon-counting error, $\sim$ 20\%, of the lightcurves and hence negligible.
Therefore, we utilize raw counts from the individual observations without
correction. 


Figure \ref{lc} shows thus obtained 33-bin lightcurves of the three sources
with time bins of 1 day.
The rms variation during five observations over two
years becomes more than 3 times as large as the average Poisson error, 
as shown in table \ref{Xrayprop}. 
To examine whether the sources are variable within each observation, we
calculated the $\chi^2$ of the lightcurves against the assumption of a
constant intensity.
While sources No.32 and No.37 exhibit statistically-significant short-term
variations, source No.28 does not.
Thus, we infer that source No.28 varies on relatively long time scales,
while the others on shorter time scales. Below, we quantify
these inferences employing our analysis method.

\section{Structure function Analysis \label{sf}}


While the lightcurves span more than 700 days, each of them contain only
33 data points. In order to estimate PSDs from these sparse lightcurves,
we utilize the ``forward method'' analysis incorporating the SF, following
Iyomto, Makishima (2001). To estimate PSDs more quantitatively, 
we improved the method in several points over the original one utilized by
these authors. First, we utilize a modified PSD model considering the Poisson
noise effect. Second, according to Timmer, Konig (1995),
we randomize not only the phase, but also the amplitude, of
each Fourier component. Third, we normalize the simulated lightcurves
referring to the PSD, rather than re-normalizing them so as to have a given
rms variability. Below we describe our procedure.

\begin{figure*}[h]
 \begin{center}
   \FigureFile(150mm,300mm){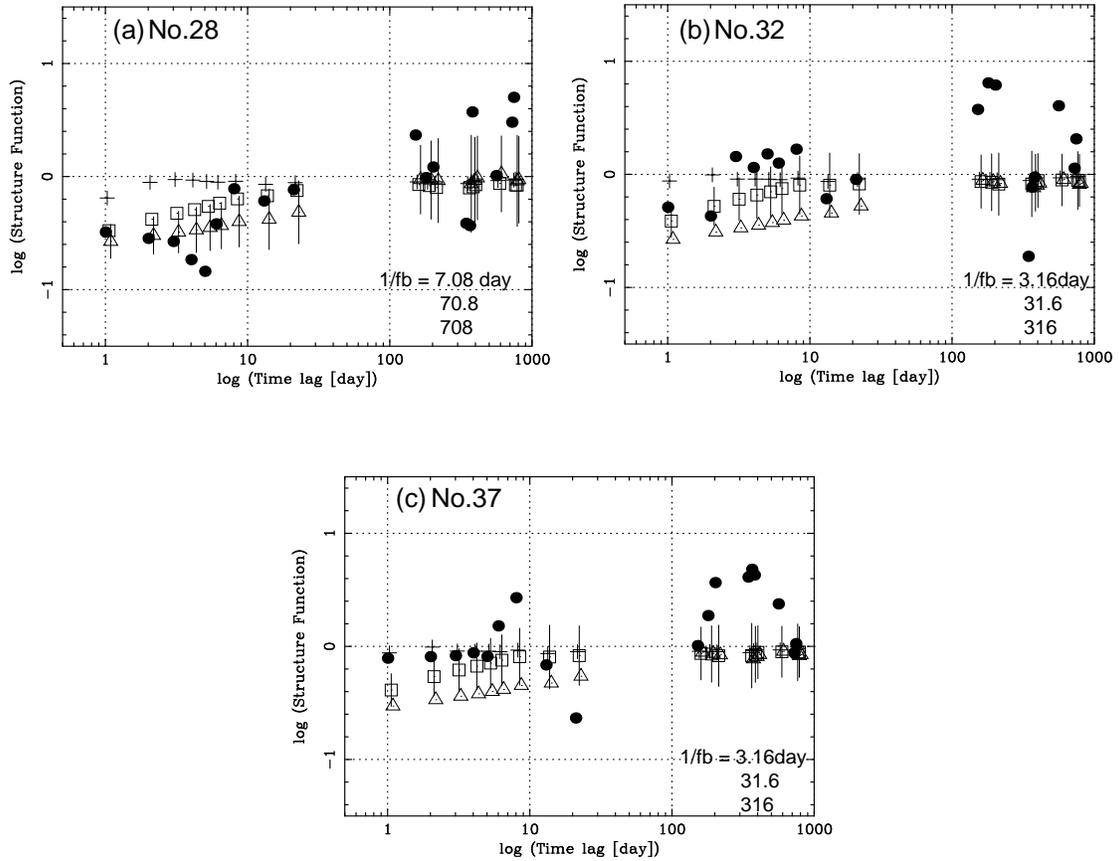}
 \end{center}
\bigskip
\caption{
         Observed binned SFs (filled circles) of the three sources,
         to be compared with the ensemble averages of simulated SFs
         after applying the window function and binning.
         An arbitrarily unit is utilized in the vertical axis.
         The simulated SFs have been calculated for three different values
         of 1/$f_{\rm b}$,
         as given in the bottom right of each panel, and specified by crosses,
         open squares, and triangles, in the order of increasing 1/$f_{\rm b}$.
         The PSD slope is fixed at $\alpha=-1.5$.
         Only for the best simulated SF, we show error bars which
         represent the standard deviation among 1000 Monte-Carlo simulations.
         \label{psd}}
\end{figure*}

We have converted the observed lightcurves (figure \ref{lc}) into SFs, as
shown in figure \ref{psd}. To suppress the scatter of the SFs especially at
large time lags, we have binned them into 18 appropriate intervals in the
time lag. All the SFs thus keep increasing monotonically as the time lag
increases. In previous studies of relatively bright sources,
the broken power-law shape, an empirical PSD of AGNs (e.g., Pounds, McHardy
1988; Edelson, Nandra 1999), was assumed as the PSD model (section 1). This
time, because of the relatively poor signal statistics, we must properly
consider the Poisson noise effect. We therefore adopt a modified PSD model of
the form (Iyomoto 1999) as  
\begin{equation}
  P(f) = P_0(f)+W,
\label{eqn:moPSD1}
\end{equation}
with 
\begin{equation}
\begin{array}{c}
     P_0(f) = \left\{
    \begin{array}{ll}
    C_0 & ( f_1<f<f_{\rm b} ) \\
    C_0 (f/f_{\rm b})^{\alpha} & ( f_{\rm b}<f<f_2 ) 
    \end{array}
    \right.\\
\label{eqn:moPSD2}
\end{array}
\end{equation}
and
\begin{equation}
W = 
\begin{array}{cc}
  \displaystyle
  W_0  \int_{f_1}^{f_2}{P_0(f)~{df}},
  \end{array}
\label{eqn:moPSD3}
\end{equation}
where $f$ denotes the frequency, $f_1$ and $f_2$ are its lower and upper
bounds respectively, $f_{\rm b}$ is the characteristic frequency called
``break frequency'', $C_0$, $W$ and $W_0$ are constants, and $\alpha$ is
the slope index. $C_0$ is determined uniquely by the rms variation of the
lightcurves, while the Poisson noise is represented by a constant term $W$,
with $W_0$ its ratio to the rms variation given in table \ref{Xrayprop}. 

Utilizing this PSD model for a given value of $f_{\rm b}$ and $\alpha$,
we simulate 1000 lightcurves, each consisting
of 4096 bins with 1-day bin width. Each lightcurve is based on different
randomization of both amplitude and phase of the Fourier components.
We adopt the total length of simulated lightcurves
$\sim$ 6 times longer than the actual data, considering that 1/$f_1$ of actual
PSD is infinite. As to $W_0$, we utilize the ratio of the Poisson noise to the
observed rms variation. We set the area of the PSD including the Poisson noise
so as to be equal to the rms variation of the observed lightcurves.
Then, considering the sampling window, we made the simulated
lightcurves sparse, and converted them into simulated SFs.
We repeated the same procedure for
1/$f_{\rm b} =10^{0.1}, 10^{0.15}, 10^{0.2}$ ...
and $10^{2.85}$ days, and $\alpha = -0.40, -0.45, -0.50$ ...
and $-2.40$.

Figure \ref{psd} shows an ensemble-average of 1000 simulated SFs,
for representative values of 1/$f_{\rm b}$, compared with the observed
SF of each source. Because we determined the normalization of the model
PSD using the rms variation and the Poisson noise of the observed lightcurves,
we directly compare the observed and simulated SFs without any further
re-normalization; this is one major improvement over Iyomoto, Makishima
(2001). We thus find, for instance, that the observed SF of source No.28
can be reproduced better by the simulation with 1/$f_{\rm b}=710$ days,
than by the other two simulations. 

\begin{figure*}[h]
 \begin{center}
     \FigureFile(150mm,300mm){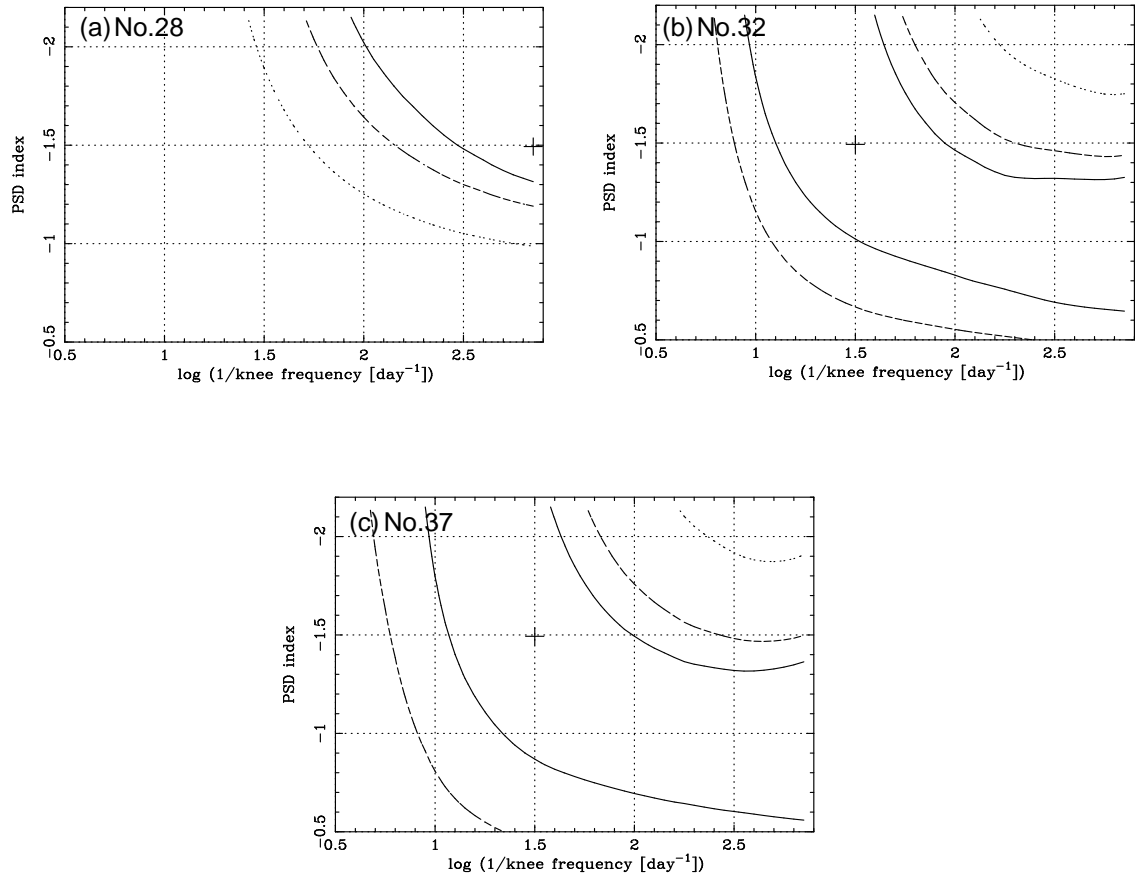}
 \end{center}
\bigskip
 \caption{
         Confidence contours of the PSD parameters of the three sources, 
         expressed on the 2-dimensional plane of $1/f_{\rm b}$ and $\alpha$. 
         Solid, dashed and dotted lines indicate 68\%, 90\% and 99\% confidence 
         regions, respectively. The cross represents the best-fit parameters when
         the PSD slope is temporally fixed at $\alpha=-1.5$ (see text). 
         The source name is shown in the top left of each panel.
         \label{cont}}
\end{figure*}

In order to compare the actual and simulated SFs more quantitatively, 
we utilized the $\chi^2$ technique referring to the dispersion of the 1000
simulated SFs as described in Iyomoto, Makishima (2001).
Figure \ref{cont} shows thus obtained 68\%, 90\% and 99\% confidence regions,
presented on the 2-dimensional plane of $\alpha$ and 1/$f_{\rm b}$.
Thus, we can constrain the values of 1/$f_{\rm b}$
as $>$ 140 days for source No.28, and 6$\sim$250 days for sources No.32
and No.37, as long as we fix $\alpha$ at $-1.5$ which is typical of AGNs
(e.g., Lawrence et al. 1987;
McHardy, Czerny 1987; Hayashida et al. 1998; Nowak, Chang 2000).
However, $\alpha$ is known to scatter nearly by $\pm$0.5 among AGNs
(Lawrence, Papadakis 1993).
Therefore, we must consider a region $-2<\alpha<-1$ in figure \ref{cont}.
Then, the constraints become much looser; 
1/$f_{\rm b}$ $>$ 50 days for source No.28, and $\gtrsim$ 6 days
for the other two sources, at a 2-sigma confidence level.
Further, if we take 3-sigma confidence level, we can place no constraints
on these two objects any longer, while 1/$f_{\rm b}>$ 25 days for source No.28.

\section{Discussion}

We have estimated the break frequency $f_{\rm b}$ of the X-ray intensity
variation of the three QSOs in the Lockman Hole field.
Allowing the PSD slope index $\alpha$ between $-2$ and $-1$ as typical
values of AGNs, 1/$f_{\rm b}$ can be constrained as $>$ 25 days for source
No.28, at a 3-sigma confidence level, 
although those of sources No.32 and No.37 are unconstrained.

We then compare the result on source No.28 with those
obtained previously on X-ray emitting BH objects including QSOs.
Because of the difficulty of long-term observations in the
X-ray band, there are few QSOs whose 1/$f_{\rm b}$ is constrained;
Fiore et al. (1998), utilizing the lightcurves obtained with ROSAT
spanning over 6 years, evaluated the PSD of PG 1440$+$356 and discovered
that it flattens below 2$\times10^{-6}$ Hz (1/$f_{\rm b}\sim$6 days). 
Using Ginga, Hayashida et al. (1998) obtained a power-law like PSD of 3C 273
between 10$^{-2}$ and 10$^{-5}$ Hz, which indicates $f_{\rm b}\leq10^{-5}$
Hz, (1/$f_{\rm b}\geq$1.2 days), if any.
Thus, our result on source No.28 is consistent with these long variation
time scales ($\gtrsim$ several days) for luminous AGNs including QSOs.

We can estimate the system size and further the BH mass of source No.28,
assuming that the variation time scale is approximately proportional to
the system size and emission is not relativistically beamed. Because the
system size reflects the Schwarzschild radius, $1/f_{\rm b}$ is considered
to be proportional to the BH mass $M_{\rm B}$. This relation may be written
as
\begin{equation}
M_{\rm BH} = 10 \times 0.1 / f_{\rm b} [{\rm Hz}] \hspace*{0.5em}M_{\odot}
\label{eqn:BHMass}
\end{equation}
where 0.1 Hz and 10 $M_{\odot}$ are the parameters of Cyg X-1 as a standard
(Makishima 1988; Miyamoto et al. 1992, 1994). 
Then, the BH mass of source No.28 falls in the range $\gtrsim 10^{7} M_{\odot}$,
in a general agreement with the mass of QSOs estimated from the reverberation
mapping method (Kaspi et al. 2000; Gu et al. 2001).

So far, we have assumed that the PSD becomes flat below $f_{\rm b}$.
However, this may not be exactly true, since the PSD of Cyg X-1 in fact
has two breaks at $\sim 1$ Hz and $\sim 0.1$ Hz (e.g., Hayashida et al.
1998; Belloni \& Hasinger 1990);
the slope index $\alpha$ changes from $\sim-2$ to $-1$ at 1 Hz, and $-1$ to
$0$ at  0.1 Hz. Therefore, below $f_{\rm b}$, the PSD of our target sources
might flatten to $\alpha \sim -1$ rather than to zero.
To examine such a case, we again performed the Monte-Carlo simulation for
the three sources, assuming that the PSD slope changes from $-1.5$ to
$-1$ at $f_{\rm b}$. We have then obtained almost the same best-fit break
frequencies, with larger errors because the change in the PSD slope is now
less conspicuous. In this case, our constraints on 1/$f_{\rm b}$ become of
even lower significance.


\vspace*{3cm}


\begin{table}[h]
 \caption{Log of ROSAT observations of the Lockman hole field.\label{obslog}}
 \begin{center}
  \begin{tabular}{rcccc}\hline\hline
  ID & ${\rm Sequence}$ & Start ${\rm Date}^{\ast}$ & End ${\rm Date}^{\ast}$ &
       ${\rm Exposure}^{\dag}$\\\hline
    1 & rp900029a00  & 91/04/16 14:32  &    05/21   08:43      &64903  \\
    2 & rp900029a01  & 91/10/25 07:37  &    11/02   18:56      &24030  \\
    3 & rp900029a02  & 92/04/15 16:18  &    04/24   08:12      &65602  \\
    4 & rp900029a03  & 92/11/29 07:41  &    11/29   11:15      &2080   \\
    5 & rp900029a04  & 93/04/26 22:09  &    05/09   07:03      &46697  \\
   \hline
  \end{tabular}
 \end{center}
$\ast$ The start and the end time of the observation,
       in year/month/day hour:minutes 
       and month/day hour:minutes, respectively.\\
$\dag$ Exposure time in seconds after the data screening.

\end{table}

\bigskip
\bigskip


\begin{table}[h]
 \caption{Properties of the selected X-ray sources. \label{Xrayprop}}
 \begin{center}
  \begin{tabular}{cccc}\hline\hline
   Source ${\rm No.}^\ast$ & 28 & 32 & 37 \\\hline

   RA (2000)$^\ast$
   & 10$^{\rm h}$  54$^{\rm m}$  21$^{\rm s}\hspace{-0.1cm}$.1
   & 10$^{\rm h}$  52$^{\rm m}$  39$^{\rm s}\hspace{-0.1cm}$.7
   & 10$^{\rm h}$  52$^{\rm m}$  47$^{\rm s}\hspace{-0.1cm}$.9\\
   
   Dec (2000)$^\dag$
   & 57$^{\circ}$ 25$^{\prime}$ 44$^{\prime\prime}\hspace{-0.15cm}$.5
   & 57$^{\circ}$ 24$^{\prime}$ 31$^{\prime\prime}\hspace{-0.15cm}$.7   
   & 57$^{\circ}$ 21$^{\prime}$ 16$^{\prime\prime}\hspace{-0.15cm}$.3 \\
    
   ${\rm Red shift}^\dag$
   & 0.205 
   & 1.113 
   & 0.467\\

   log ${L_x}^\ast$
   & 43.59 & 44.69 & 43.46\\

   AGN type$^\dag$
   & II & I & I\\

   Radio flux$^\ddag$
   & 0.80 & 0.14 & --\\

   count ${\rm rate}^\S$
   & 1.19 ${\pm0.26}$
   & 2.19 ${\pm0.34}$
   & 1.22 ${\pm0.27}$\\

   Poisson Error ratio (\%)$^\|$
   & 30
   & 28
   & 34\\\hline

  \end{tabular}
 \end{center}
$\ast$ Referring to Paper I. $L_x$ corresponds to
       the 0.5--2.0 keV luminosity.\\
$\dag$ Referring to Papers II and III.\\
$\ddag$ The total flux density (in units of mJy) at 1.5 GHz, 
        referring to Paper IV.\\
$\S$   The background-subtracted 0.1--2.4 keV PSPC count rate
       (in units of $10^{-2}$ cts s$^{-1}$)
       averaged over the five observations.
       Errors represent the average 1$\sigma$ Poisson error
       of 1-day binned lightcurves.\\
$\|$ The ratio of the average Poisson noise
       to the rms variation over five observations. See text.\\
\end{table}



\begin{figure*}[h]
 \begin{center}
   \FigureFile(150mm,300mm){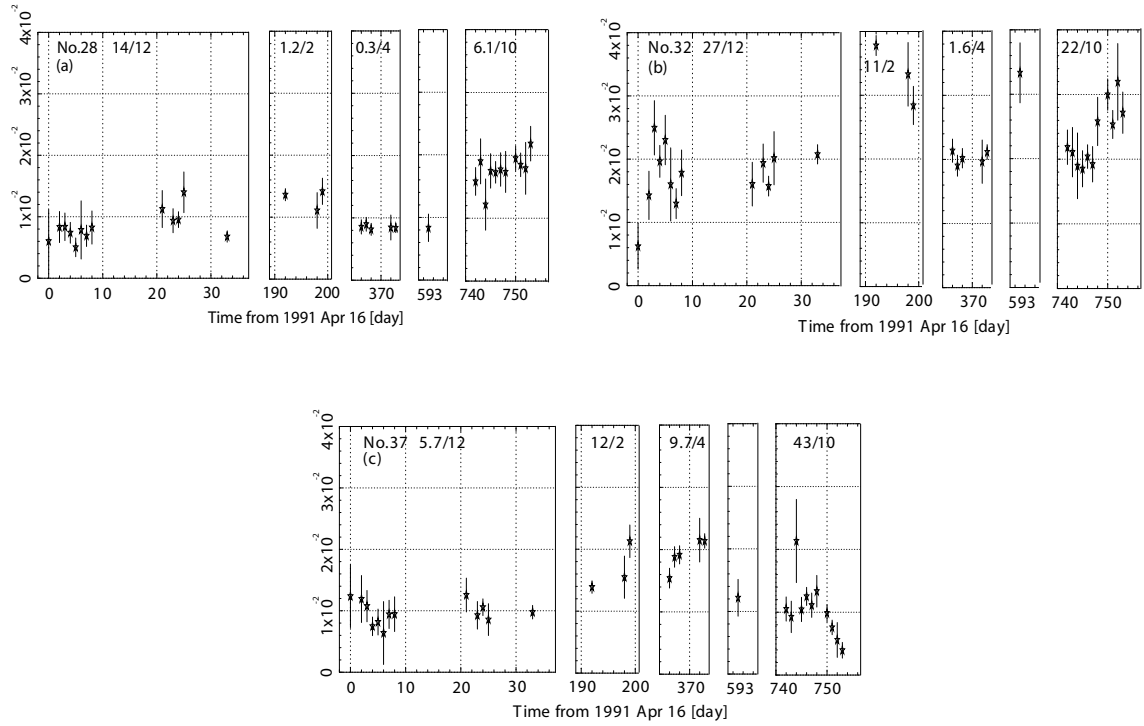}
 \end{center}
\bigskip
\caption{
         The lightcurves of the three sources with 1-day bin width.
         The vertical axis represents the 0.1--2.4 keV flux
         (erg cm$^{-2}$ s$^{-1}$)
         after subtracting the background. Error bars represent
         the 1$\sigma$ Poisson error.
         Individual panels correspond to the lightcurves from the first 
         to the fifth observations.
         The source name is shown in the top left of each panel, together
         with the $\chi^2$/d.o.f. values
         against the hypothesis of constant intensity.
         Note the difference in the flux scale among sources.
         \label{lc}}
\end{figure*}


\end{document}